\documentclass[twocolumn,amsmath,amssymb,prl]{revtex4-1}

\usepackage{graphicx}
\usepackage{dcolumn}
\usepackage{bm}

\begin{document}



\title{Direct measurements of the dynamical correlation length indicate its divergence at an athermal glass transition}

\author{Z. Rotman}
\author{E. Eisenberg}

\affiliation{Raymond and Beverly Sackler School of Physics and Astronomy,
Tel Aviv University, Tel Aviv 69978, Israel}



\begin{abstract}
The super-cooled $N3$ model exhibits an increasingly slow dynamics as density approaches the model's random closest packing density. Here, we present a direct measurement of the dynamical correlation function $G_4(r,t)$, showing the emergence of a growing length scale $\xi_4$ across which the dynamics is correlated. The correlation length measured, up to $120$ lattice sites, power-law diverges as the density approaches $\rho_t$, the density at which the fluid phase of the model is predicted to terminate. It is shown that the four-point susceptibility, often used as an agent to estimate $\xi_4$, does not depend simply on the latter. Rather, it depends strongly on the short-range behavior of $G_4(r,t)$. Consequently, $\chi_4$ peaks before $\xi_4$ reaches its maximal value. The two quantities should therefore be studied independently.
\end{abstract}


\maketitle

Dynamics of super-cooled liquid glass-formers is characterized by
non-exponential temporal relaxation (see recent review \cite{cavagna}). Two natural, but fundamentally different, explanations can be put forward. Relaxation might be locally exponential, but the typical relaxation timescale varies spatially. Global response functions become non-exponential upon spatial averaging, due to the spatial distribution of relaxation times. Alternatively, relaxation might be complicated and inherently non-exponential, even locally. Experimental and theoretical works \cite{ediger} suggest that while the latter mechanism is likely to contribute, relaxation (at times of order of the global relaxation time or shorter) is indeed spatially heterogeneous. At an ideal glass transition, the relaxation time diverges and the dynamics remains heterogeneous for arbitrary long times. The study of dynamical heterogeneities has attracted much interest recently \cite{bennemann,garrahan,merolle,berthier1,toninelli,stein,karmakar}, as part of the attempt to decipher the mystery of glassy dynamics.

A physical characterization of dynamic heterogeneity entails the determination of the typical lifetime of the heterogeneities, as well as their typical lengthscale.
A clear demonstration of the heterogeneous dynamics is obtained by tracking the mobile particles at a given time. One observes clusters of highly mobile particles as well as clusters of particles barely moving at all \cite{vidal,keys,candelier}. The typical size of these clusters is quantified by the spatial decay of the mobility correlation function. One therefore looks for the correlation between the displacements over a time interval $t$ of particles at mutual distance $r$. This mobility-mobility correlation function was first introduced in \cite{lancaster}, as a tool to discover cooperative regions
in numerical simulations of glass-forming liquids. However, the first attempts to analyze these correlation function \cite{lancaster} were very limited, and thus showed no increasing correlation length. It turns out that the essential ingredient to find the dynamical correlations is not really the mobility, but rather the fact that we are calculating a four-point correlation function, in contrast with standard two-point functions. The four point density correlation
\begin{eqnarray}
\label{g4} G_4(r,t)&=&\langle\rho(0,0)\rho(0,t)\rho(r,0)\rho(r,t)\rangle  \nonumber \\
& &-\langle\rho(0,0)\rho(0,t)\rangle\langle\rho(r,0)\rho(r,t)\rangle ,
\end{eqnarray}
was first suggested in \cite{dasgupta} in order to look
for a growing correlation length. Yet, direct measurements of $G_4$ turn out to be technically demanding. Instead, it is common to study its spatial integral, which is nothing but the fluctuations of the two-point correlation function $C(t)=\langle\rho(0,t)\rho(r,t)\rangle-\langle\rho(0,t)\rangle\langle\rho(r,t)\rangle$, or the four-susceptibility
\begin{equation}
\label{chi4} \chi_4(t)=\int G_4(r,t)dr=N[\langle C(t)^2\rangle - \langle C(t)\rangle^2].
\end{equation}
The four-susceptibility was indeed shown to exhibit an appreciable increase with the waiting time \cite{parisi-c}, which was interpreted as a sign for an increase of the dynamical correlation length characterizing the decay of $G_4(r,t)$. The four-point susceptibility was later used extensively to measure dynamical heterogeneities approaching the glass transition and to imply the existence of growing correlation length. The time-dependence of $\chi_4(t)$ was suggested to indicate the mechanism underlying the transition \cite{toninelli}.

Various theories lead to different relations between the correlation length $\xi_4$, characterizing the spatial decay of $G_4(r,t)$, and $\chi_4$ \cite{biroli1,biroli2}. Thus, direct measurements of dynamical correlations can play a vital role in distinguishing the various theoretical approaches. In fact, it was pointed out in \cite{doliwa,toninelli} that $\chi_4$ and $\xi_4$ might depend differently on time. However the calculation of dynamic correlations is hard especially in numerical simulations, because very large systems are needed in order to determine $\xi_4$ unambiguously.
Previous measurement of dynamic correlations were performed indirectly by either scaling the binder cumulant at $\tau_4$ (time at which $\chi_4(t)$ peaks) \cite{karmakar} or fitting the low $q$ behavior of $G_4(q,t)$ \cite{toninelli,stein}. These measurements have shown moderate correlation lengths, up to $10$ inter-particle distances. Clearly, these do not describe very well the behavior at a regime in which correlation lengths (are expected to) diverge. Here we report measurements of dynamical correlation based on the evaluation of the four-point density correlation $G_4(r,t)$ throughout the relaxation process. Having measured $G_4(r,t)$ directly, the correlation lengths is then extracted by fitting an exponentially decaying function to the long range behavior of $G_4(r,t)$.

\begin{figure}
\includegraphics[width=6.9cm,height=6.9cm,angle=-90]{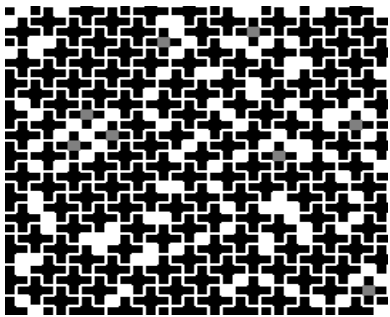}
\caption{(Color Online) Random packing of the N3 model as produced by the cooling protocol, density set to 0.17. The few particles that are movable at the end of the cooling phase are marked by a gray spot at their center.}
\label{fig-N3}
\end{figure}

\begin{figure}
\includegraphics[width=8cm,height=8cm,angle=-90]{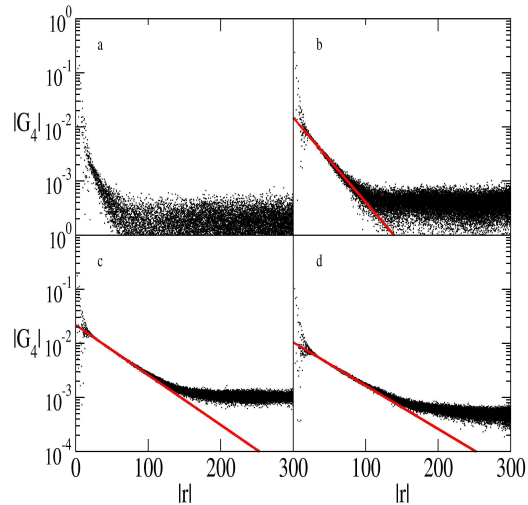}
\caption{(Color Online) Measurements of $G_4(r,t)$ during dynamical relaxation of the N3 model, at density $\rho=0.1713$. (a) t=5000 (b) t=10000, a clear exponentially regime emerges. (c) t=20000, the correlation length further increases. (d) t=30000, further increase of the correlation length. Note that the amplitude of the correlation function $G_4(r,t)$ is smaller in this case, leading to an overall smaller integral $\chi_4$ as compared to the time presented in (c). The exponential fit in red may be used to extrapolate $G_4(r,t)$ to larger distances (see text).}
\label{fig-g4fit}
\end{figure}

\begin{figure}
\includegraphics[width=6.9cm,height=6.9cm,angle=-90]{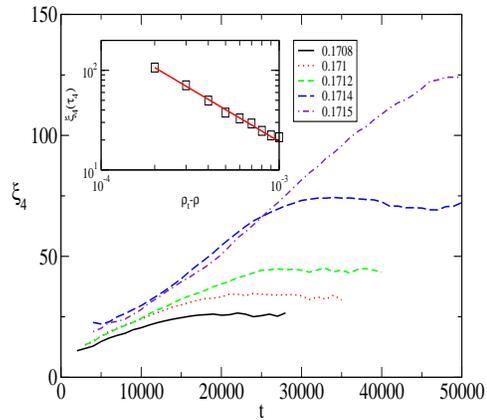}
\caption{(Color Online) Time dependence of the dynamical correlation length for several values of the density. The correlation length increases with time, and then saturates for as long as we are able to measure it. The correlation length as measured at $\tau_4$ power-law diverges with critical exponent of 1.0 (inset).}
\label{fig-xi}
\end{figure}

The $N3$ model is a simple 2D model on a square lattice. Particles are cross-shaped pentamers, interacting only through hard-core exclusion which blocks up to the 3$^{\rm rd}$ nearest neighbor (see Fig. \ref{fig-N3}). The model is known to undergo a first order solidification transition \cite{bellemans,orban,eliasher}, where density jumps from $\rho_f\simeq0.161 $ to $\rho_s\simeq0.191$ \cite{eliasher} (the closest packing density is $0.2$). The behavior of this model in the supercooled fluid regime was analyzed using the $R$ matrix method \cite{baramr,us} based on the Mayer cluster coefficients of N3. The analysis predicted a critical termination of the supercooled fluid phase at non trivial density $\rho_t=0.1717$, close to the random closest packing density of this model \cite{N3-rcp}. Extensive MC simulations revealed that the dynamic relaxation of the supercooled fluid becomes increasingly slow as the density increases, diverging exactly at the density predicted by $R$ matrix analysis \cite{N3}. It is therefore plausible to relate the dynamical arrest in the $N3$ model to the predicted termination of the thermodynamic equation of state predicted by the $R$ matrix. The $N3$ system is then a simple and convenient model-system for studies of slow dynamics in quenched deeply supercooled fluids.

Here we employ the advantage of the the simple two dimensional N3 system and study in detail the dynamic correlation length near the
predicted termination of the fluid supercooled phase. Simulations of the N3 model allowed for an easy and direct measurements of $G_4(r,t)$ throughout the whole relaxation process for a wide range of system sizes. We performed Monte Carlo simulations of the model following the protocol presented in \cite{N3}: simulation starts with an infinitely fast cooling, where particles are added whenever possible and diffuse otherwise, this process is stopped when the desired density is reached. The system is then left to relax diffusively at a fixed density, and measurements are performed during this relaxation process. Figure \ref{fig-g4fit} presents the behavior exhibited at the different time regimes. We have studied lattices of sizes up to $1200\times 1200$ sites in the current study (up to $2000\times 2000$ sites in the past). Simulations of such large systems are crucial, as significant finite-size effects are shown to persist to large lattice sizes \cite{N3}. In concordance, we measure correlation lengths as large as $120$ lattice-sites ($50$ inter-particle distances). The correlation length is estimated by fitting an exponential form to the measured $G_4(r,t)$. The only freedom in the fit is the spatial range upon which to fit the exponential form: in all of the $\xi_4$ measurements below we used the range $30<r<90$, even though for cases with large $\xi_4$ the fit captures the behavior far beyond this chosen range, as seen in Figure \ref{fig-g4fit}).

The picture that emerges from these simulations is the following: at short times $G_4(r,t)$ decays at short range, with no exponential tail observed. However, on intermediate times one clearly sees an exponential decay regime following the short-range decrease. At later times, (later then $\tau_4$), one again sees only the short range behavior.
Figure \ref{fig-xi} presents the time dependence of the dynamic correlation length $\xi_4$, for various densities of the N3 model.
We find that the correlation length $\xi_4(t)$ grows with time, approaching a plateau. It is not possible to measure $\xi_4$ at very long times, where $G_4(r,t)$ is very small beyond the short-range regime, and it is not at all clear whether there is an exponential regime in these long times. However, for as long as we are able to measure $\xi_4$ it does not seem to increase or decrease. Furthermore,
the time dependence of $\chi_4(t)$ is determined by the short range correlation (which dictates the amplitude of the long range behavior) as well as the correlation length itself. Due to the dependence of both $\xi_4$ and the amplitude of $G_4(r,t)$ on the time $t$, the time $\tau_4$ at which $\chi_4$ peaks differs from the time at which $\xi_4$ reaches its maximal value. These findings highlights the non-trivial relation between $\chi_4$ and $\xi_4$, and the need to study both quantities separately. In addition, our measurements show that the dynamically correlated regions have compact (and not fractal) form in all times measured.

\begin{figure}
\includegraphics[width=6.9cm,height=6.9cm,angle=0]{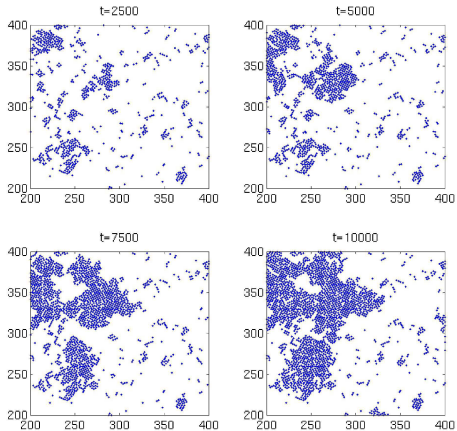}
\caption{(Color Online) Displaying relaxation dynamics of N3 model, for $\rho=0.1713$. Dots represent mobile particles, the growth of mobile regions as well as no creation of new mobile regions with time is observed. An area of 200x200 was chosen out of simulated system of 1000x1000 sites }
\label{fig-move}
\end{figure}

Looking at the microscopic structure of the relaxation dynamics in the N3 model, one observes a clear heterogeneous picture similar to that seen, for example, in the kinetically constrained triangular lattice gas models (1)-TLG and (2)-TLG \cite{pan}. The dynamics can be described by a growth of mobile regions in the system, as seen in Figure \ref{fig-move}, until all of the particles moved when $C(t)=0$. In course of time, the mobile regions, or clusters, grow, but there is no creation of new clusters. The measured correlation length fits this observed dynamics. The increase in the correlation length corresponds to the growth of clusters of mobile particles, while in later times further cluster growth is blocked by the neighboring clusters, resulting in a saturation of correlation length.

Measurements of $G_4$ have revealed an additional phenomenon. As seen in Figure \ref{fig-g4fit}, $G_4(r,t)$ does not decay to 0 at very long ranges, of order of the system's size. Close examination of this unexpected feature shows that the (unphysical) infinite-range correlation observed depends on system size, decreasing like $1/L^2$. It thus follows that its contribution to $\chi_4$, the spatial integral over $G_4$, persists for very large systems.
It turns out that the source of these spurious infinite-range correlations is the initial configuration (that is, correlations introduced during the cooling protocol). To show this, we looked at the correlation function of the density $\rho_m(r,t=0)$ of particles which are free to move at $t=0$ (end of cooling phase and start of the measurement time). The global density of movable particles at time $t=0$ is just $m = \langle \rho_m(r,t=0)\rangle$, and the generalized susceptibility for this density is given by
$$\chi_m=V[\langle m^2\rangle-\langle m\rangle^2]$$. This quantity is closely related to $\chi_4$ at short times, since at these short times particles that indeed moved are, by and large, those that were free to move at $t=0$. Looking at the correlation function of the movable particles density $G_m(r)=\langle m(0)m(r)\rangle-\langle m\rangle^2$ one observes the same infinite-range contributions. It thus follows that infinite-range correlations in the density of the movable particles emerge during the cooling protocol we used. Similar infinite-ranged correlations have been observed for other out-of-equilibrium systems \cite{derrida}. This global effect is then amplified in the course of dynamics, where movable particles grow into clusters, and affects strongly dynamical measurements of $\chi_4$. The effect is washed out only for relatively long times, often of order $\tau_4$ and more. Obviously, these infinite-range contributions to $\chi_4$ have nothing to do with the dynamics in general, or with the dynamical correlation length in particular.


\begin{figure}
\includegraphics[width=6.9cm,height=6.9cm,angle=-90]{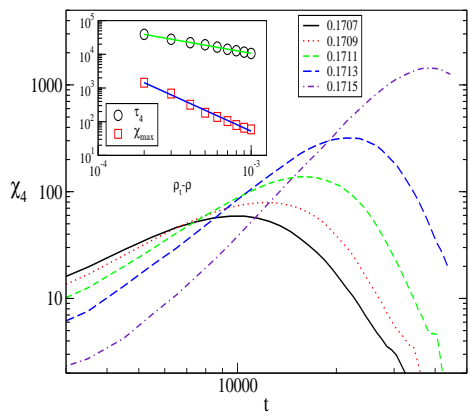}
\caption{(Color Online) Measurements of $\chi_4(t)$ corrected for removal of global effects created while cooling the system. Maximum value as well as the time of maximum power law diverges (inset) as the termination density $\rho_t$ is approached. Maximum value diverges with exponent of 2.0 and the the time $\tau_4$ with exponent of 0.8}
\label{fig-chi4}
\end{figure}

Similar effects are expected in other studies, as long as the system is not equilibrated before the measurement starts. Such equilibration is always desired, but usually not possible when studying systems that are prone to crystallization (such as hard-spheres, or the N3 model). Thus, direct measurements of $\chi_4$ in such cases are prone to be affected by the effect described above. However, if one measures $G_4(r,t)$, a correction scheme for the initial condition is possible. The value of $\chi_4$, often calculated using $N \cdot Var(C(t))$, does contain global contributions which are artifacts of the cooling protocol. Based on the observed dynamics, we claim that only the growth of the mobile regions is a local dynamical property that should be taken into account, and the infinite-range contribution (present even at $t=0$ and therefore not connected to the dynamics) should be removed. We therefore use the measured dynamic correlations $G_4(r,t)$ in order to correct the $\chi_4$ measurements: the exponential decaying region of $G_4(r,t)$ is extrapolated to infinite $r$, and $\chi_4$ is calculated by integrating $G_4(r,t)$ exactly over the short-range and adding the contribution of the integral over the exponential regime for larger distances.

The results for the dynamic $\chi_4$ are presented in Figure \ref{fig-chi4}. Peak values of $\chi_4$ diverges with a power law as $\rho_t$ is approached $\chi_{4,max}\sim(\rho_t-\rho)^{-2.0}$. The time-dependence of $\chi_4(t)$ underscores its dependence on the short range behavior of $G_4(r,t)$ rather than the long exponential tail. The short range correlation determines the coefficient multiplying the exponential decay, and has stronger time dependence then $\xi_4(t)$ and so determines $\chi_4(t)$ time dependence.

In summary, we present here a direct measurement of the dynamical correlation function $G_4(r,t)$, showing the emergence of a growing length scale $\xi_4$ across which the dynamics is correlated. One observes a significant increase in the correlation length, up to $~120$ lattice sites. As density approaches $\rho_t$, the density at which the fluid phase of the model is predicted to terminate, the correlation length $\xi_4$ power-law diverges with $(\rho_t-\rho)$.  In concordance, peak values of $\chi_4(t)$  diverge, suggesting that the slow-down of relaxation processes near the termination of the fluid branch results from a growth in the dynamical correlations. Yet, we show that $\chi_4$ by itself cannot be used as a reliable indirect measure of $\xi_4$ time-dependence.
\begin{acknowledgements}
We are grateful to Ludovic Berthier for important discussions and insightful comments.
\end{acknowledgements}

\end{document}